\pdfoutput=1
\documentclass[useAMS,usenatbib]{mn2e}
\usepackage{graphicx}
\usepackage{times}
\usepackage{wasysym}
\usepackage{url}
\usepackage{lineno}
\usepackage{amssymb}
\usepackage{epstopdf}

\newcommand{\HMS}[3]{$#1^{\mathrm{h}}#2^{\mathrm{m}}#3^{\mathrm{s}}$}
\newcommand{\DMS}[3]{$#1^\circ #2\arcmin #3\arcsec$}
\newcommand{\hessj}{HESS~J1640$-$465}
\newcommand{\pwn}{XMMU~J164045.4$-$463131}
\newcommand{\fermi}{2FGL~1640.5$-$4633}
\newcommand{\rxj}{RX~J1713.7$-$3946}
\newcommand{\snr}{G338.3$-$0.0}
\newcommand{\g}{$\gamma$}

\newcommand{\UNITS}[1]{\,\mathrm{#1}}

\title[An exceptionally luminous TeV \g-ray SNR]{\hessj\ -- an
  exceptionally luminous TeV \g-ray supernova remnant}

\author[H.E.S.S. Collaboration]{\normalsize H.E.S.S. Collaboration,
 A.~Abramowski,$^{1}$
 F.~Aharonian,$^{2,3,4}$
 F.~Ait Benkhali,$^{2}$
 A.G.~Akhperjanian,$^{5,4}$
 E.~Ang\"uner,$^{6}$
 G.~Anton,$^{7}$
 \newauthor \normalsize
 S.~Balenderan,$^{8}$
 A.~Balzer,$^{9,10}$
% ...
\\
$^1$
Universit\"at Hamburg, Institut f\"ur Experimentalphysik, Luruper Chaussee 149, D 22761 Hamburg, Germany\\
$^2$
Max-Planck-Institut f\"ur Kernphysik, P.O. Box 103980, D 69029 Heidelberg, Germany\\
$^3$
Dublin Institute for Advanced Studies, 31 Fitzwilliam Place, Dublin 2, Ireland\\
$^4$
National Academy of Sciences of the Republic of Armenia, Yerevan\\
$^5$
Yerevan Physics Institute, 2 Alikhanian Brothers St., 375036 Yerevan, Armenia\\
$^6$
Institut f\"ur Physik, Humboldt-Universit\"at zu Berlin, Newtonstr. 15, D 12489 Berlin, Germany\\
$^7$
Universit\"at Erlangen-N\"urnberg, Physikalisches Institut, Erwin-Rommel-Str. 1, D 91058 Erlangen, Germany\\
$^8$
University of Durham, Department of Physics, South Road, Durham DH1 3LE, U.K.\\
$^9$
DESY, D-15738 Zeuthen, Germany\\
$^{10}$
Institut f\"ur Physik und Astronomie, Universit\"at Potsdam,  Karl-Liebknecht-Strasse 24/25, D 14476 Potsdam, Germany\\
% ...
}

\begin{document}

\date{Accepted 2014 January 17. Received 2014 January 17; in original
  form 2013 October 23}
\pagerange{\pageref{firstpage}--\pageref{lastpage}} \pubyear{2014}

\maketitle

\label{firstpage}

\begin{abstract}
  The results of follow-up observations of the TeV $\gamma$-ray source
  \hessj\ from 2004 to 2011 with the High Energy Stereoscopic System
  (H.E.S.S.) are reported in this work. The spectrum is well described
  by an exponential cut-off power law with photon index $\Gamma=2.11
  \pm 0.09_{\mathrm{stat}} \pm 0.10_{\mathrm{sys}}$, and a cut-off
  energy of $E_c = 6.0^{+2.0}_{-1.2}$\,TeV. The TeV emission is
  significantly extended and overlaps with the north-western part of
  the shell of the SNR \snr. The new H.E.S.S. results, a re-analysis
  of archival \emph{XMM-Newton} data, and multi-wavelength
  observations suggest that a significant part of the \g-ray emission
  from \hessj\ originates in the SNR shell. In a hadronic scenario, as
  suggested by the smooth connection of the GeV and TeV spectra, the
  product of total proton energy and mean target density could be as
  high as $W_p n_H \sim 4 \times 10^{52}(d/10
  \mathrm{kpc})^2$\,erg\,cm$^{-3}$.
\end{abstract}

\begin{keywords}
  radiation mechanisms: non-thermal, ISM: supernova remnants, ISM:
  individual objects: \snr
\end{keywords}

\maketitle

\section{Introduction}\label{sec:intro}

Starting in 2004 the Galactic Plane Survey \citep{HESS:GPS06}
performed by the H.E.S.S. Collaboration, using an array of imaging
atmospheric Cherenkov telescopes (IACTs), led to the discovery of
nearly 70 new sources in the very-high-energy (VHE, E$>$100\,GeV)
\g-ray regime \citep{Carrigan13}. The challenge since then has been to
associate these sources with astrophysical objects seen in other
wavelengths and to identify the underlying radiation mechanisms. A
large fraction of the Galactic VHE \g-ray population could be
associated with regions with recent star-forming activity and to
objects at late stages of stellar evolution such as supernova remnants
(SNRs) and the nebulae produced by powerful young pulsars \citep[for a
review, see e.g.][]{HintonHofmann09}. In many cases where an
astrophysical counterpart to the VHE \g-ray emission could be
identified, however, the nature of the underlying particle population
remains unclear. Highly energetic \g-ray emission could be either
produced by relativistic electrons or protons (and heavier
nuclei). Relativistic hadrons undergo inelastic scattering with nuclei
in the interstellar medium (ISM), producing $\pi^0$-decay \g-ray
emission. Ultra-relativistic electrons, on the other hand, can
up-scatter low-energy photons present at the acceleration site via the
Inverse Compton (IC) process. In very dense media Bremsstrahlung
losses of electrons can significantly contribute to the generated
\g-ray emission. IACTs can play a key role in identifying the
underlying particle population and studying non-thermal processes in
\g-ray sources by localising the emission region and constraining the
energy spectrum at very high energies.

The VHE \g-ray source \hessj\ was discovered by H.E.S.S. in the
Galactic Plane Survey \citep{HESS:GPS06} and is positionally
coincident with the SNR \snr\ \citep{Whiteoak96}. Using
\emph{XMM-Newton} observations \citet{1640:Funk} detected a highly
absorbed extended X-ray source (\pwn) close to the geometric centre of
the SNR and within the H.E.S.S. source region. The X-ray and VHE
\g-ray emission components were interpreted as synchrotron and IC
emission from relativistic electrons in a pulsar wind nebula
(PWN). Observations with \emph{Chandra} confirmed the presence of the
extended nebula and identified a point-like source which was suggested
to be the associated pulsar \citep{Lemiere09}. Recently,
\citet{1640:Castelletti11} analysed new high-resolution
multi-frequency radio data of \snr\ but could only set upper limits on
the radio flux from a potential extended radio
nebula. \emph{Fermi}-LAT observations revealed a high-energy (HE,
100\,MeV$<$E$<$100\,GeV) \g-ray source coincident with \hessj\
\citep{1640:Slane10}, also designated \fermi\ in the two-year
\emph{Fermi}-LAT catalogue \citep{Fermi:2yr}. Note that no pulsation
has been found in any wavelength band so far. Due to the large \g-ray
to X-ray ratio luminosity \citep[$L_\gamma/L_{X} \simeq
30$;][]{1640:Funk}, \citet{1640:Slane10} inferred an evolved PWN with
a low magnetic field and an injection spectrum that consists of a
Maxwellian electron population with a power-law tail \citep[as
e.g. proposed by][]{Spitkovsky08} to reproduce the broadband spectral
energy distribution (SED) in a leptonic PWN scenario. A hadronic
origin of the \g-ray emission was considered to be unlikely as it
would require rather high ambient densities ($n\simeq100$\,cm$^{-3}$),
implying intense thermal radiation in X-rays from the SNR shell that
has so far not been detected.

\citet{Lemiere09} performed a detailed study of the gaseous
environment of \snr, and based on the HI absorption features, derived
a distance of $(8-13)$\,kpc. A recent study of the nearby stellar
cluster Mercer~81 and the giant HII region G338.4+0.1 by
\citet{Davies12} supports this estimate, which implies that \hessj\ is
the most luminous VHE \g-ray source known in the Galaxy. Throughout
this work, a distance of 10\,kpc is assumed. Since the original
discovery of \hessj, the available H.E.S.S. exposure towards this
source has quadrupled w.r.t the data used in \citep{HESS:GPS06}, and
advanced analysis methods are now available that allow for a much more
detailed spectral and morphological study of the VHE \g-ray
emission. In this work, H.E.S.S. follow-up studies and a re-analysis
of \emph{XMM-Newton} data are presented. Both the broadband SED and
the TeV morphology reveal evidence for proton acceleration in the SNR
shell of \snr.

\section{H.E.S.S. observations and results}\label{sec:hess}
H.E.S.S. is an array of five imaging atmospheric Cherenkov telescopes
located in Namibia designed to detect VHE \g-rays. The fifth telescope
started operation in September 2012. All H.E.S.S. data used to perform
the studies described below have been taken between May 2004 and
September 2011 with the four-telescope array \citep{HESS:Crab}. The
total dead time corrected live time amounts to 63.4\,hr, compared to
14.3\,hr in the original publication \citep{HESS:GPS06}. Observations
have been performed at zenith angles between $20^\circ$ and $65^\circ$
with a mean value of $\sim33^\circ$. The data were recorded with
pointing offsets between $0.2^\circ$ and $1.8^\circ$ with a mean value
of $1.1^\circ$ from the \hessj\ position. Data were analysed using a
standard Hillas-type H.E.S.S. analysis\footnote{The software package
  HAP version 12-03-pl02 with version32 of the lookup tables was
  used.} for the event reconstruction and a boosted decision tree
based event classification algorithm to discriminate $\gamma$-rays
from the charged particle background \citep{TMVA}. All results were
cross-checked by an independent analysis and calibration for
consistency \citep{Model++}.

\subsection{Morphology}
\begin{figure}
\begin{center}
  \resizebox{\hsize}{!}{\includegraphics[]{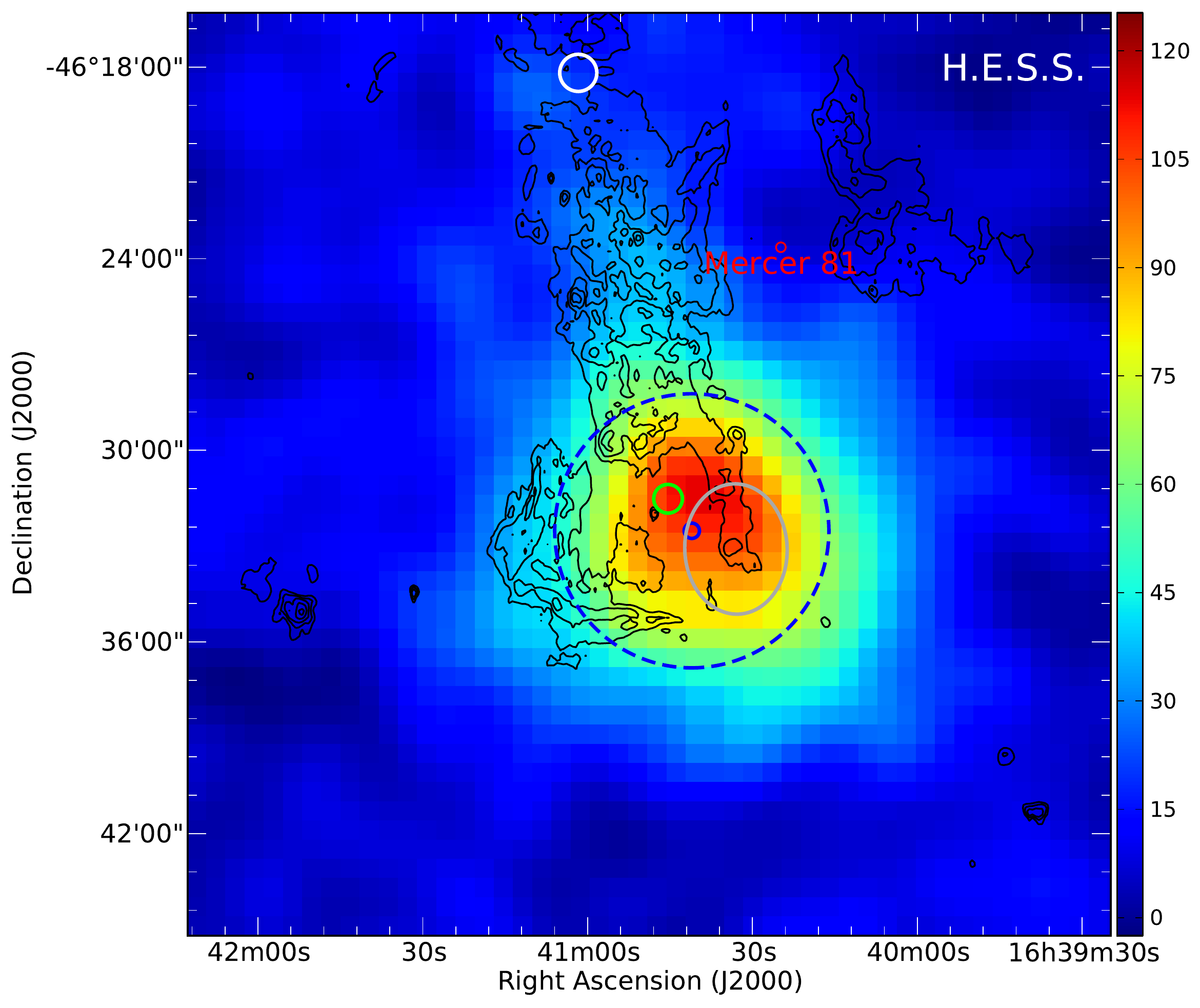}}
  \caption{H.E.S.S. excess map smoothed with a 2D Gaussian with
    $0.017^{\circ}$ variance and the best-fit position (statistical
    errors only) and intrinsic Gaussian width overlaid as blue solid
    and dashed lines. 610\,MHz radio contours are shown in black
    \citep{1640:Castelletti11}. The green circle indicates the
    position of the candidate PWN \pwn, and in gray the best-fit
    position of the Fermi source \fermi\ is given. The white circle
    indicates the source HESS~J1641$-$463 \citep{Oya2013} and the
    region of high radio emission connecting \hessj\ and
    HESS~J1641$-$463 is the HII region G338.4+0.1. The progenitor of
    \snr\ is potentially associated with the massive young stellar
    cluster Mercer~81 \citep{Davies12}.}
\label{fig1}
\end{center}
\end{figure}

The source position and morphology have been obtained with \emph{hard}
cuts and using the ring background estimation method
\citep{HESS:Background}. In this setup a minimum intensity in the
camera image of 160\,p.e. is required, resulting in an energy
threshold of $E_{\mathrm{th}} = 600$\,GeV and a point spread function
(PSF) with 68\% containment radius of $r_{68} = 0.09^\circ$ for the
morphology studies. The fit of a symmetric two-dimensional Gaussian
profile, convolved with the H.E.S.S. PSF with \emph{Sherpa}
\citep{Sherpa01} gives a best-fit position of RA \HMS{16}{40}{41.0}
$\pm~1.0^{\mathrm{s}}_{\rm stat} \pm~1.3^{\mathrm{s}}_{\rm sys}$ and
Dec \DMS{-46}{32}{31} $\pm~14\arcsec_{\rm stat} \pm~20\arcsec_{\rm
  sys}$ (J2000), consistent with the previously published value
\citep{HESS:GPS06}. The systematic error on the best-fit position
originates from the pointing precision of the H.E.S.S. array of about
$20\arcsec$. The source is intrinsically extended with a Gaussian
width of $\sigma_S = (4.3 \pm 0.2)\arcmin$. This extension is
$1.6\arcmin$ ($\sim$$2\sigma$) larger than in the original
publication, which can be understood as fainter emission belonging to
\hessj\ that can now be revealed with the increased data
set. Figure~\ref{fig1} shows the H.E.S.S. best-fit position and
extension overlaid on the VHE \g-ray excess map. The VHE \g-ray source
encloses the northern part of the SNR shell of \snr, the candidate PWN
\pwn\ \citep{1640:Funk} and the Fermi-LAT source \fermi\
\citep{1640:Slane10,Fermi:2yr}. Figure~\ref{fig1} also shows some
indication for an asymmetric extension of the emission along the
northern part of the shell and towards the newly discovered source
HESS~J1641$-$463 \citep{Oya2013}. This extension is also seen as
residual VHE \g-ray emission when subtracting the source model from
the sky map, indicating that the symmetric Gaussian model for \hessj\
is an oversimplification. The residual emission could indicate some
emission in between \hessj\ and HESS~J1641$-$463. This component is
however not detected with high significance, making a discussion of
its origin difficult in this context. Morphological fits in energy
bands do not reveal any significant change in best-fit position and/or
extension, which would have indicated a change in source morphology
with energy \citep[as e.g. seen in the PWNe HESS~J1825$-$137 or
HESS~J1303$-$631;][]{HESS:1825_Edep, HESS:1303}.

\subsection{Spectrum}\label{sec:spectrum}

\begin{figure} \centering
  \resizebox{\hsize}{!}{\includegraphics[]{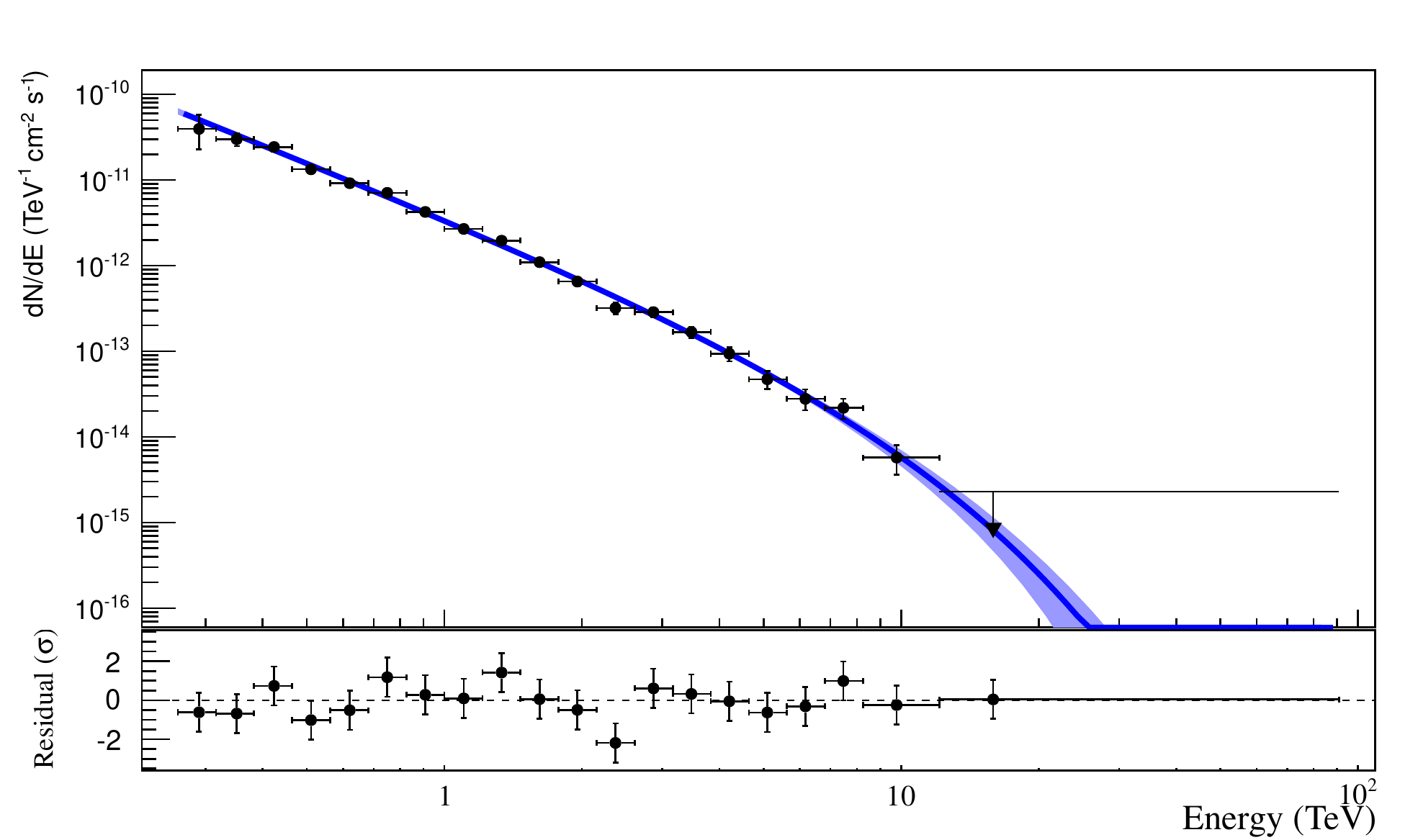}}
  \caption{VHE \g-ray spectrum of \hessj\ (top) and flux residuals
    (bottom) extracted within the 90\% containment radius (see text).
    Also shown is the best-fit power law, plus exponential cut-off
    model and 68\% error band. All spectral points have a minimum
    significance of 2$\sigma$. The last point is the differential flux
    upper limit in this energy band at 95\% confidence level.}
\label{fig:spectrum}
\end{figure}
 
The VHE \g-ray spectrum is shown in Figure~\ref{fig:spectrum}, and has
been extracted using \emph{std} cuts (60\,p.e. minimum image
intensity, $E_{\mathrm{th}} = 260$\,GeV), using the reflected region
background method \citep{HESS:Background} and forward folding with a
maximum likelihood optimisation \citep{Piron01} from the 90\%
containment radius of the VHE \g-ray emission of \hessj\ of
$0.18^\circ$ around the best-fit position. The fit of a power law with
exponential cut-off:
$dN/dE=\Phi_0\times(E/1~\mathrm{TeV})^{-\Gamma}e^{-E/E_c}$ results in
a photon index $\Gamma=2.11 \pm 0.09_{\mathrm{stat}} \pm
0.10_{\mathrm{sys}}$, a differential flux normalisation at 1\,TeV of
$\Phi_0$ = $(3.3 \pm 0.1_{\mathrm{stat}} \pm 0.6_{\mathrm{sys}})
\times~10^{-12} \UNITS{TeV^{-1}\,cm^{-2}\,s^{-1}}$ and a cut-off
energy of $E_c = 6.0^{+2.0}_{-1.2}$\,TeV. The systematic errors on
flux norm and index for this data set are based on the difference seen
between the main and cross-check analysis and are a result of
uncertainties in e.g. atmospheric conditions, simulations, broken
pixels, analysis cuts, or the run-selection. The fit probability $p$
for an exponential cut-off power law model is $p \sim 36\%$, whereas
the fit probability for a pure power law model is $p \sim 1\%$. The
luminosity of \hessj\ above 1\,TeV at 10\,kpc distance is
$L_{>1\mathrm{TeV}} \simeq 4.6\times 10^{35} (d/10\,{\rm
  kpc})^2$\,erg\,s$^{-1}$, a factor of $\sim 2.8$ higher than that of
the Crab nebula.

\begin{figure} \centering
  \resizebox{\hsize}{!}{\includegraphics[]{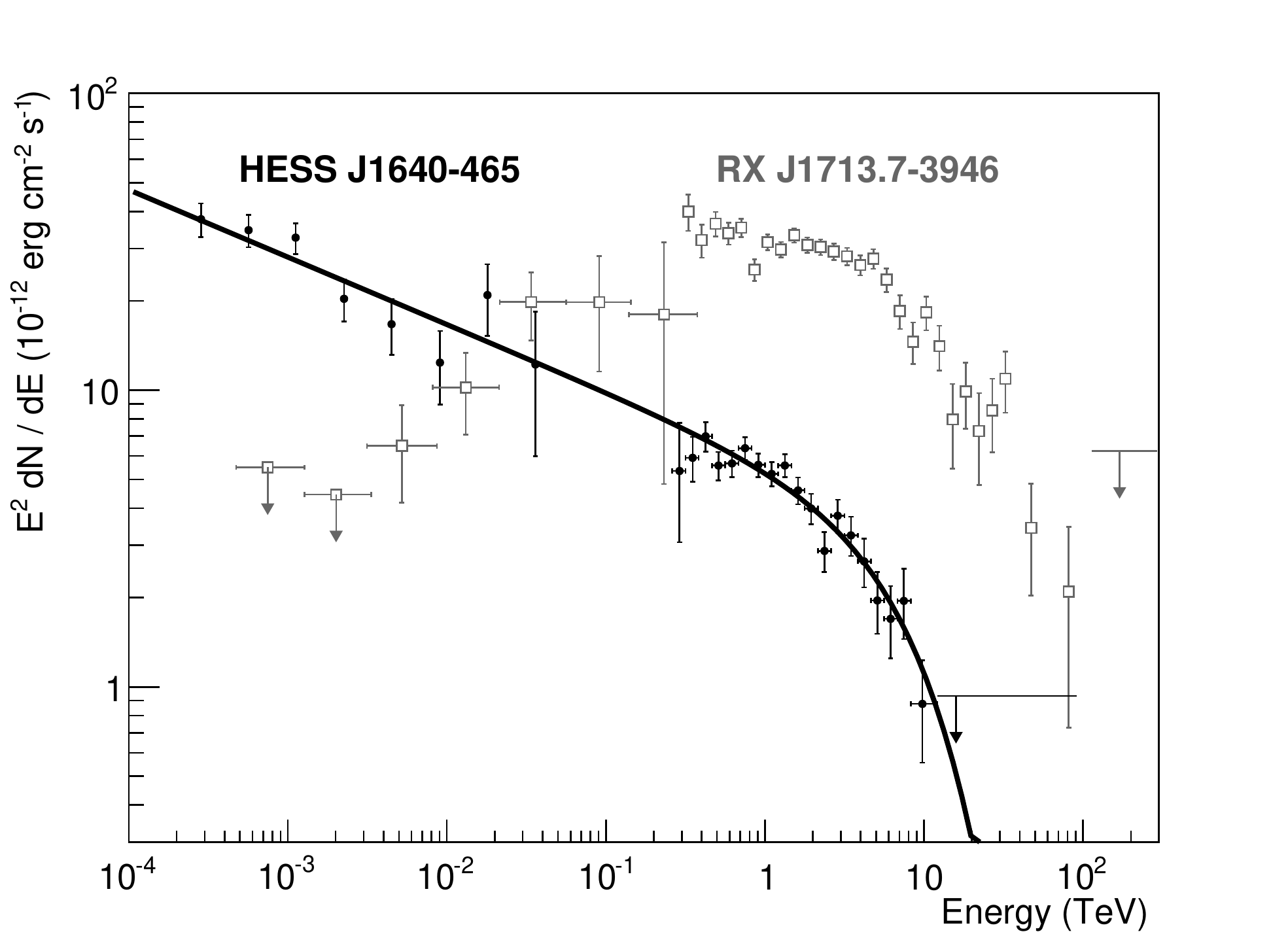}}
  \caption{Comparison of the HE and VHE \g-ray spectra of \hessj\
    (filled circles) and \rxj\ (open squares). Data for \rxj\ are from
    \citet{Fermi:1713} and \citet{HESS:1713_cor}, GeV data of \hessj\
    is from \citet{1640:Slane10}. Also shown is the best-fit
    exponential cut-off power law model to the full \g-ray spectrum
    (Table~\ref{tab:spectra}).}
\label{fig:rxj1713}
\end{figure}

The photon index as reconstructed with the new H.E.S.S. data at TeV
energies is compatible with the photon index as reconstructed in the
GeV domain \citep{1640:Slane10, Fermi:2yr, Fermi:1FHL}. A simultaneous
exponential cut-off power law fit to the GeV data points as derived by
\citet{1640:Slane10}, and new TeV data between 200\, MeV and 90\, TeV
(shown in Figure~\ref{fig:rxj1713}) has been performed. The result of
this fit is summarised in Table~\ref{tab:spectra} and shows that the
flux at 1\,TeV, the photon index as well as the cut-off energy are
consistent with the fit to the H.E.S.S.-only data. The fit has a
$\chi^2$ of 21 for 24 degrees of freedom (\emph{d.o.f.}) with a
probability of 63\%\footnote{The fit has been performed on the binned
  H.E.S.S spectrum shown in Figure~\ref{fig:spectrum} and on the GeV
  spectrum from \citet{1640:Slane10} taking into account statistical
  errors only.} and implies that no break in the \g-ray spectrum
between the \emph{Fermi} and H.E.S.S. energy range is required in
order to describe the data. 

\begin{table*}
  \centering
  \caption{Best-fit spectrum results of the new H.E.S.S. data as shown
    in Figure~\ref{fig:spectrum}, and in combination with the GeV spectrum
    from \citet{1640:Slane10}.}
  \begin{tabular}{cccccc}
    \hline
    Data & $E_{\rm{min}}$ & $E_{\rm{max}}$ & $\Gamma$ & $\Phi_0$ &
    $E_{c}$ \\
    & & & & $10^{-12}$\,cm$^{-2}$\,s$^{-1}$ & TeV \\\hline
    H.E.S.S. & 260\,GeV & 90\,TeV & $2.11 \pm 0.09$ & $3.3 \pm 0.1$ & $6.0^{+2.0}_{-1.2}$\\
    H.E.S.S. + \emph{Fermi}-LAT & 200\,MeV & 90\,TeV & $2.23 \pm 0.01$ & $3.7 \pm 0.2$ & $8.8^{+2.3}_{-1.5}$ \\ \hline
  \end{tabular}
  \label{tab:spectra}
\end{table*}

\section{XMM-Newton data analysis}\label{sec:xmm}

\citet{1640:Funk} reported the detection of the candidate PWN \pwn\
with \emph{XMM-Newton} and introduced it as a potential counterpart of
\hessj.  As becomes clear from Fig.~\ref{fig1} the VHE \g-ray emission
region also overlaps with the northern part of the shell of SNR \snr.
To investigate \g-ray emission scenarios related to the SNR, the
\emph{XMM-Newton} data (ObsID: 0302560201) were re-analysed to derive
an upper limit for diffuse X-ray emission originating from the
northern part of the shell. For the analysis the Science Analysis
System (SAS) version 12.0.1 was used, supported by tools from the
FTOOLS package and XSPEC version 12.5.0 \citep{1996ASPC..101...17A}
for spectral modelling. The data are affected by long periods of
strong background flaring activity resulting in net exposures of only
$5.9$\,ks (PN) and $13.5$\,ks (MOS), following
the suggested standard criteria for good-time-interval filtering. To
detect and remove point-like X-ray sources the standard
\emph{XMM-Newton} SAS maximum likelihood source detection algorithm
was used in four energy bands ($(0.5-1.0)$\,keV, $(1.0-2.0)$\,keV,
$(2.0-4.5)$\,keV, and $(4.5-10.0)$\,keV). Events around all sources
detected in any of these bands were removed from a region
corresponding to the 95\% containment radius of the \emph{XMM-Newton}
PSF at the respective source position in the detector. The total flux
upper limit was derived assuming that the remaining count-rate from a
polygon region enclosing the northern part of the shell is due to
background. A power-law model with photon index $\Gamma_X = -2$ was
applied to constrain non-thermal leptonic emission. Two different
absorption column densities as found in the literature,
$N_\mathrm{H,1}$ = 6.1$\times$10$^{22}$\,cm$^{-2}$ \citep{1640:Funk}
and $N_\mathrm{H,2}$ = 1.4$\times$10$^{23}$\,cm$^{-2}$
\citep{Lemiere09}, have been considered. No diffuse X-ray emission
coincident with the SNR shell was detected with this data set. The
resulting 99\% confidence upper limits for the unabsorbed flux
($(2-10)$\,keV) are $F_{99} (N_\mathrm{H,1}) = 4.4 \times
10^{-13}$\,erg\,cm$^{-2}$\,s$^{-1}$ and $F_{99} (N_\mathrm{H,2}) =
8.3\times10^{-13}$\,erg\,cm$^{-2}$\,s$^{-1}$. These values have been
scaled up by 11\% to account for the missing area due to excluded
point-like sources.

\section{Discussion}\label{sec:discussion}

The H.E.S.S. source encloses the PWN candidate \pwn\ as well as the
north-western half of the incomplete shell of \snr. The comprehensive
multi-wavelength data available together with the new H.E.S.S. and
\emph{XMM-Newton} results allow for a much more detailed investigation
of the SED and hence the underlying non-thermal processes to be
carried out. As the evolutionary state of \snr\ is essential for the
discussion, the age of the SNR is estimated, and the environment in
which it likely expanded is investigated. These estimates will form
the basis for the discussion of the origin of the non-thermal emission
in a PWN and SNR scenario.

\subsection{Age and Environment of \snr}\label{sec:age}

The age and environment of the SNR have a large influence on the
interpretation and modeling of the emission scenario and thus deserve
discussion in this context. Previous estimates put the age of the SNR
in the range of $(5-8)$\,kyr \citep{1640:Slane10}, however, as becomes
evident from the discussion below, it may be significantly younger
than that.

If the X-ray PWN is indeed related to the SNR, then \snr\ originated
from a core-collapse supernova explosion of a massive star. Such stars
usually modify the surrounding medium through strong stellar winds,
creating a cavity of relatively low density surrounded by a
high-density shell of swept-up material.
\citep[see][]{Weaver77,Chevalier99}. Such a wind-blown bubble scenario
has never been considered for this object, but needs to be explored
for a detailed discussion of the \g-ray emission mechanisms possibly
at work in \hessj. These cavities have significant impact on the
evolution of the subsequent supernova shock front, and such scenarios
have been evoked to explain the properties of other SNRs like the
Cygnus Loop \citep[e.g.][]{Levenson98}, RCW\,86 \citep{Vink97}, and
\rxj\ \citep{Fukui03}, all of which have physical diameters similar to
\snr.  \citet{Chevalier99} estimated the size of wind-blown cavities
by requiring a pressure equilibrium between the inside of the bubble,
which has been pressurised by the total energy of the wind:
$1/2\dot{M}v^2_\mathrm{w}\tau$, and the surrounding medium. Here,
$\dot{M}$ is the mean mass-loss rate, $v_\mathrm{w}$ is the wind speed
and $\tau$ is the lifetime of the star. With a distance of 10\,kpc,
the radius of the observed shell of \snr\ is 10\,pc, which is assumed
here to be comparable to the size of the wind-blown bubble.  Such
sizes can be achieved by a typical $\sim$20\,$M_\odot$ O-type star
with $\tau\simeq$\,7\,Myr, $\dot{M} \simeq
10^{-7}\,M_\odot$\,yr$^{-1}$, and $v_\mathrm{w}\simeq
2600$\,km\,s$^{-1}$, evolving in an HII region with temperature 10\,kK
\citep{Osterbrock1989} and average density of $n\sim$~150\,cm$^{-3}$
\citep[see below,][]{Kudritzki2000,Muijres2012}. This corresponds to a
total mass loss in the main sequence phase of 0.7\,$M_\odot$. An
extreme case that may provide a lower limit to the age of the SNR can
be derived by the assumption that the remaining material inside the
cavity solely originates from the stellar wind. The mean number
density then is $n_0\sim$~0.01\,cm$^{-3}$ with a total mass swept up
by the SNR shock of 0.7\,$M_\odot$. This means that the SNR shock
would evolve freely expanding up to the radius of the wind-blown
bubble. Assuming average shock velocities between
$(5000-10000)$\,km\,s$^{-1}$ the age of the SNR would be $(1-2)$\,kyr,
which is considerably younger than the estimate of $(5-8)$\,kyr by
\citet{1640:Slane10}, owing to the lower density.

In addition to the SNR age, also the density of the ISM in the
immediate vicinity of the shock region has major impact on the
interpretation of the emission scenario. The density in the shell
surrounding the wind-blown bubble can be estimated with various
methods, i.e. via thermal radio emission, thermal X-ray measurements
and HI absorption studies. \citet{1640:Castelletti11} found evidence
for thermal radioemission in the SNR shell indicating the presence of
dense material. The authors infer electron densities based on the
free-free absorption feature in the radio spectrum of ${n}_e \sim
(100-165)$\,cm$^{-3}$. No diffuse X-ray emission from the SNR shell
have been reported in \citet{1640:Funk}, and in the previous section
upper limits have been derived. \citet{1640:Slane10} argue that
therefore high gas densities are not supported. However, the lack of
observed thermal X-ray emission might be consistent with the very
large distance and high column densities inferred from the
\emph{XMM-Newton} and \emph{Chandra} spectra \citep{Lemiere09} of the
PWN \pwn ; especially if the plasma temperature is below 1\,keV. Only
for higher temperatures, as e.g. observed from Kes\,32 \citep{Vink04},
could observable thermal X-rays be expected from this
source. Particularly, SNRs evolving rapidly inside low-density
wind-blown cavities are not expected to produce significant thermal
X-ray emission. Only when the SNR shock hits the surrounding shell,
the medium in the shock region thermalises rapidly and cools extremely
fast, which makes the SNR an efficient emitter of hard thermal X-rays,
but only during a short time. Later, the temperatures are expected to
drop significantly below 1\,keV due to the decreased shock speeds of
only a few 100\,km\,s$^{-1}$ \citep[see e.g.][]{Tenorio1991}. As
outlined above, due to the high absorption towards \snr\ such emission
is not expected to be detectable.

Finally, the HI absorption feature can be used to infer a maximum
(neutral) gas density. Assuming that all of the HI gas as studied by
\citet{Lemiere09} between $-65$\,km\,s$^{-1}$ and $-55$\,km\,s$^{-1}$
is associated with \snr\ and located in a shell with 4\,pc thickness
(as supported by radio observations) at 10\,kpc, a maximum density of
$n_{H,\mathrm{max}} \simeq 600$\,cm$^{-3}$ can be derived. 
However, since some of the absorbing gas may not be associated with \snr, average neutral gas
densities $\bar{n}_H$ lower than that are also plausible. From the HI
absorption measurements and the thermal radio emission, the hydrogen
gas (neutral plus ionised) in the region is consistent with densities
of $\bar{n}_{H} \gtrsim (100 - 150)$\,cm$^{-3}$. \citet{Purcell12}
performed a survey for high-density gas ($n \gtrsim 10^4\,$cm$^{-3}$)
in NH$_3$ transition lines in the Galactic plane. With the sensitivity
of this survey and given that no emission in these transition lines is
seen towards \hessj\, a molecular cloud more massive than $\sim
8000\,M_\odot$ is not supported by the data. However, this does not
exclude the existence of smaller, similarly dense clumps of material
in the shell region (see below). There is also no maser emission
detected towards the TeV emission, which would have indicated the
interaction of a shock wave with dense material
\citep[e.g.][]{Walsh2011}.

\subsection{PWN scenario}\label{sec:pwn}

The positional coincidence of \hessj\ and \fermi\ with the candidate
X-ray PWN \pwn\ is seen as evidence for leptonic \g-ray emission from
a PWN \citep{1640:Funk, Lemiere09, 1640:Slane10}. In these scenarios,
electrons are accelerated to energies of hundreds of TeV in the PWN,
radiate via synchrotron and IC processes and produce the observed
X-ray and HE and/or VHE \g-ray emission. In the following the PWN
interpretation will be confronted with the new spectral and
morphological H.E.S.S. results and the available multi-wavelength
information.

The \g-ray spectrum of middle-aged and old PWNe is characterised by a
break in the SED of $\Delta\Gamma = 0.5$ at the energy where the
IC/synchrotron loss time of the parent electron population is similar
to the age of the source \citep[e.g.][]{HintonHofmann09}. For young
PWNe ($t \simeq 1$\,kyr) the \g-ray spectrum from interactions of
electrons with magnetic and radiation fields is effectively uncooled
up to the cut-off energy as IC and synchrotron loss times are much
longer in a typical PWN environment. This leads to a peak in the IC
and synchrotron spectra at energies just below the cut-off energy in
the electron spectrum. An IC peak (or spectral break) is seen for all
of the GeV and TeV identified PWNe \citep[e.g.][]{Grondin11,
  HESS:1825_Edep, Fermi:MSH15-52, HESS:MSH15-52, HESS:VelaX2}, but not
for \hessj. To reproduce the observed \g-ray spectral index
$\Gamma_\gamma \simeq 2.2$ for a young object ($\lesssim 2.5$\,kyr),
the injection spectrum has to be $\Gamma_e = 3.4$, as $\Gamma_e =
(2\Gamma_\gamma -1)$ -- an index significantly steeper than predicted
by Fermi acceleration theory. \citet{1640:Slane10} suggested an
additional Maxwellian low-energy electron component in order to
explain the smooth connection of the HE and VHE \g-ray spectra. As
shown in Section~\ref{sec:spectrum} the new high-quality
H.E.S.S. spectrum connects with the GeV spectrum without any
discernable features and thus does not require such a contribution.
In fact, a $\chi^2$ test of the \citet{1640:Slane10} model on the
binned GeV and TeV spectrum results in a $\chi^2 = 189$ for 25
\emph{d.o.f.} with very low probability, not supporting a significant
contribution of such a Maxwellian component. This can be compared to
the exponential cut-off power law model as shown in
Table~\ref{tab:spectra}, which has a $\chi^2 = 21$ for 24
\emph{d.o.f.}

From a theoretical point of view, the extent of the PWN is expected to
be smaller than its associated SNR \citep[e.g.][]{Blondin01}. This
prediction is supported by observations of several PWNe, including
MSH~15$-$52 \citep[][]{HESS:MSH15-52} and Vela~X
\citep{HESS:VelaX2}. The intrinsic size of \hessj\ at TeV energies,
however, is larger than \snr\ and features significant overlap with
the shell of the SNR -- a behaviour that is not seen for any other
PWN.

At radio wavelengths, \citet{1640:Castelletti11} derived upper limits
on the possible radio emission from the PWN at various wavelengths,
with the most constraining limit of
3.7$\times$$10^{-17}$\,erg\,cm$^{-2}$\,s$^{-1}$ at 610\,MHz within the
X-ray PWN. Due to the different cooling times of the underlying
electron population, the PWN is expected to have a larger extent in
radio than in X-rays \citep[e.g.][]{GaenslerSlane2006}. As no radio
emission has been detected at the X-ray PWN location, it is hard to
estimate the size and hence total flux from a potential radio PWN. The
610\,MHz map shows a deficit of emission at the X-ray PWN location and
some enhancement inside the rest of the SNR. This could be associated
with projected SNR emission, or with a relic radio PWN. For young PWNe
the peak of the radio emission is expected to be close to the pulsar
position. Since the radio surface brightness around the putative
pulsar is much lower compared to the rest of the SNR interior, this
would imply that the radio excess is related to projected shell
emission. For older systems, however, the radio PWN can very well fill
the full interior of the SNR shell. As a compromise, the limit as
given by \citet{1640:Castelletti11} is scaled up by a factor of 16 to
cover the interior of the SNR shell. In this case the radio limit is a
factor of $\sim$five below the model curves in \citet{Lemiere09} and
\citet{1640:Slane10}, and would imply a low-energy cut-off of the
underlying electron spectrum significantly higher than the 50\,GeV as
used by \citet{Lemiere09}.

In summary, the interpretation of the GeV and TeV emission as solely
originating from a PWN is very difficult as neither the \g-ray
spectrum, nor the morphology or the radio data support such a
picture. A possible solution would be that the GeV emission has a
different origin than the TeV emission. This, however, requires
fine-tuning to explain the smooth \emph{Fermi} and H.E.S.S. spectrum
and the positional coincidence of the GeV and TeV sources. Also the
TeV spectrum alone does not show any significant deviation from a pure
power law below the cut-off energy, which would be expected for a
young PWN. In fact, the radio upper limit in
\citet{1640:Castelletti11}, the X-ray data and a non-dominant IC
component in the \g-ray regime would be consistent with \pwn\ being a
young PWN \citep[c.f. Fig.~5 in][]{1640:Funk}. In general, the
featureless \g-ray spectrum over almost six decades in energy is
challenging for any leptonic model as spectral breaks and sharp
cut-offs are expected in the resulting SED due to cooling and
Klein-Nishina effects, respectively \citep[e.g.][]{HintonHofmann09}.

The TeV emission also significantly overlaps with the north-western
part of the shell of \snr\ and it is hence quite natural to explore an
origin of the non-thermal emission in the SNR shell. Especially the
spectral characteristics of \hessj\ are similar to that of prominent
Galactic SNRs interacting with molecular clouds such as W28, W51C or
IC\,443 \citep[see][and references therein]{Ohm:SB12}. In the
following the focus will be on an origin of the non-thermal emission
in the SNR shell, bearing in mind that some fraction of the total TeV
emission could plausibly originate from the PWN.

\subsection{SNR scenario}\label{sec:snr}

Given the spectral and morphological similarity of \hessj\ with other
Galactic SNRs interacting with molecular clouds, an SNR origin of the
non-thermal emission is studied in the following. In a hadronic
$\gamma$-ray emission scenario, a high-density is required to provide
sufficient target material for the relativistic protons to produce
neutral pions which subsequently decay into energetic photons
\citep[see e.g.][]{Aharonian1994}. This high density material outside
the SNR shock could either be the wind shell surrounding the stellar
wind bubble, or the dense material known to exist in the vicinity of
\hessj. The relatively low ISM density Inside the wind-blown bubble
would not be sufficient to account for the bulk of the observed
$\gamma$-ray emission, and thus the target material must be of
different origin. In the environment of \snr\ there could be at least
two possibilities for the occurrence of sufficiently dense ISM: a) As
discussed in Section \ref{sec:age} and following \citet{Chevalier99},
wind-blown bubbles are surrounded by a thin dense shell containing the
bulk of the material swept-up by the stellar wind. If the expanding
shock of \snr\ is now close to this region, accelerated protons might
interact with this dense material and subsequently produce the
observed $\gamma$ rays. b) A second possibility is that the SNR shock
expands into a highly inhomogeneous ISM towards the nearby HII region
featuring dense clumps of molecular gas surrounded by regions of
comparatively low density. Here, the particles could be efficiently
accelerated within the inter-clump medium while energetic protons can
penetrate into the dense clumps and produce the observed $\gamma$-ray
emission. This scenario has already been proposed for the young
($\sim$2\,kyr) VHE $\gamma$-ray emitting SNR \rxj\
\citep[see][]{Zirakashvili10} where dense molecular cloud cores have
been detected in the shock region \citep[e.g.][]{Sano10}.  Such ISM
conditions are probably also present in the vicinity of \snr, due to
its vicinity to a massive and dense HII region, making this emission
scenario also viable for \hessj.

In contrast to middle-aged interacting SNRs like IC\,443
\citep{Fermi:IC443} and W\,44 \citep{2010Sci...327.1103A} where the
$\gamma$-ray spectra are strongly peaked at GeV energies, \rxj\ and
other young SNRs emit a large fraction of their high-energy emission
in the TeV regime, either due to a different radiation process or
their earlier stage in evolution. Figure~\ref{fig:rxj1713} shows a
comparison between the GeV--TeV spectra of \hessj\ and \rxj\ as seen
by \emph{Fermi} and H.E.S.S. Interestingly, their spectral shapes in
the TeV regime are very similar, which could support an age younger
than $(10-20)$\,kyr for \snr. However, the GeV spectrum becomes much
harder for \rxj\ but keeps the same slope for \hessj. Leptonic models
giving rise to the observed shape of the $\gamma$-ray spectrum of
\rxj\ have been discussed in the literature quite extensively
\citep[see e.g.][]{Fermi:1713,2011ApJ...735..120Y}. However, following
\citet{Zirakashvili10}, the change in slope towards lower energies for
\rxj\ could also be explained in a hadronic scenario by the smaller
penetration depths into the dense molecular cloud cores for protons
with lower energies \citep[see also][]{Inoue12}. These particles
therefore cannot interact with the same amount of material as protons
with higher energies, giving rise to an under-luminous and harder GeV
$\gamma$-ray spectrum. The fact that this feature is not seen for
\hessj\ might indicate an older remnant than e.g. \rxj\
(i.e. $\gtrsim$~2.5\,kyr) or different diffusion properties of the
local ISM that allow also low-energy protons to fully penetrate the
dense molecular clumps. An age of 2.5\,kyr would imply some mixing of
the stellar-wind material and the ISM leading to average densities in
the wind bubble of $n_0\sim$0.1\,cm$^{-3}$
(c.f. Section~\ref{sec:age}).

When comparing the TeV morphology of \hessj\ to \snr\ (Fig.~\ref{fig1}) it
becomes clear that $\gamma$-ray emission only shows significant
overlap with the north-western (NW) part of the radio shell. Thus, in
a hadronic scenario the lack of emission from the south-eastern (SE)
shell needs to be explained. In such a model the $\gamma$-ray emission
is expected to follow the distribution and the density of available
target material in the shock region. Indeed, a correlation between the
molecular and atomic gas and the VHE $\gamma$-ray intensity from \rxj\
has recently been reported by \citet{Fukui12}. Thus, if dense target
material is much more abundant in the northern region of \snr\
compared to the south, the observed TeV morphology of \hessj\ is
consistent with a hadronic scenario.
\begin{figure} \centering
  \resizebox{\hsize}{!}{\includegraphics[]{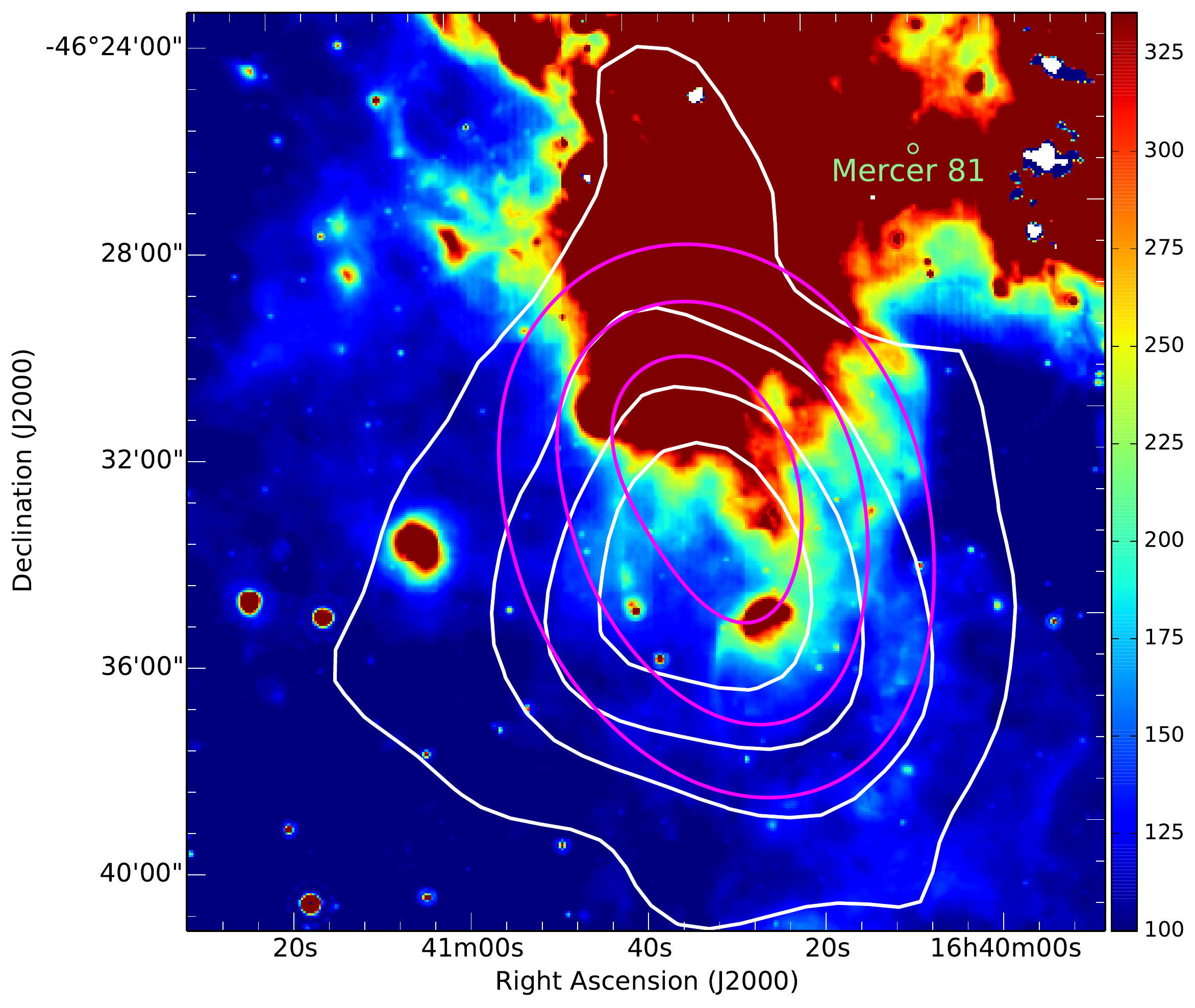}}
  \caption{\emph{Spitzer} MIPS 24\,$\mu$m image in units of MJy
    sr$^{-1}$ with overlaid contours from the smoothed H.E.S.S. excess
    map (white) and contours of the north-western part of the SNR
    shell from the 610\,MHz image, convolved with the H.E.S.S. PSF
    (magenta, c.f. Fig.~\ref{fig1}).}
\label{fig:spitzer}
\end{figure}
Figure~\ref{fig:spitzer} shows the \emph{Spitzer} MIPS
\citep{Rieke2004} 24\,$\mu$m image of this region, which essentially
traces the abundance of interstellar dust and dense HII star-forming
regions. Here it can be seen that the mean infrared intensity towards
the NW part is a factor of $\sim$5 higher than towards the SE area of
the shell. Therefore, the different densities could indeed give rise
to the observed morphology. To further test the hypothesis of the NW
shell being the origin of the VHE $\gamma$-ray emission, only this
part of the radio shell was used as a template and convolved with the
H.E.S.S. PSF. The resulting contours are over-plotted on the
\emph{Spitzer} image in Fig.~\ref{fig:spitzer} and show a good
agreement with the VHE \g-ray excess contours from H.E.S.S.

\begin{figure} \centering
  \resizebox{\hsize}{!}{\includegraphics[]{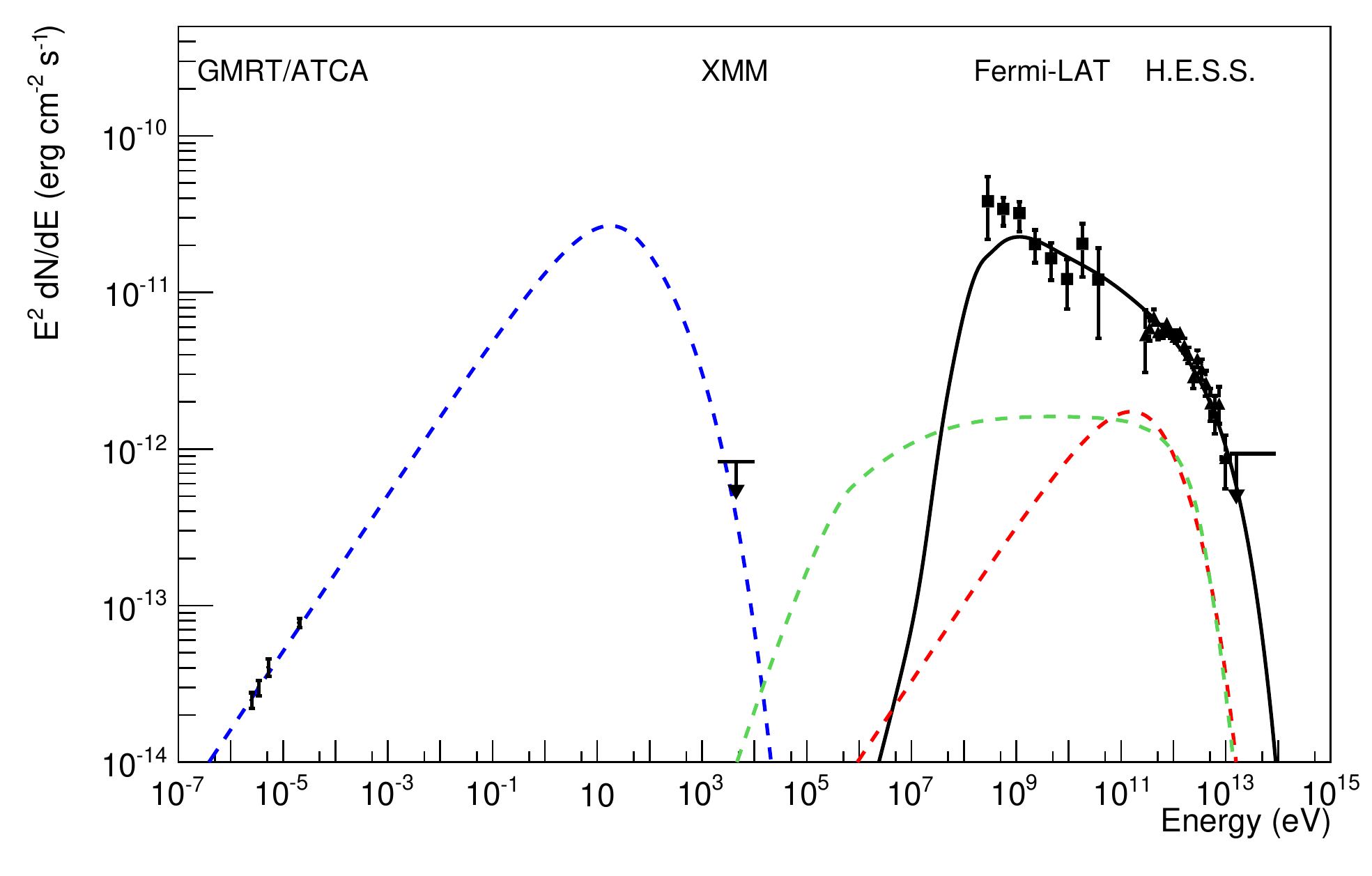}}
  \caption{HE and VHE \g-ray spectrum of \hessj\ as given in
    \citet{1640:Slane10} and shown in Figure~\ref{fig:spectrum},
    respectively. The X-ray limit has been derived in the northern
    part of the radio shell and assuming the higher column density as
    derived by \citet{Lemiere09} (see Figure~\ref{fig1} and text), and
    the radio data is from \citet{1640:Castelletti11}, scaled by a
    factor of 0.5, assuming that half of the radio emission comes from
    the northern part of the shell. The long-dashed blue and red
    dash-dotted curves are synchrotron and IC emission from
    non-thermal electrons, respectively. The green dashed curve is the
    Bremsstrahlung component and the solid black curve is hadronic
    $\pi^0$-decay \g-ray emission.}
\label{fig:snr}
\end{figure}

Figure~\ref{fig:snr} shows the measured SED of \snr\ along with the
new H.E.S.S. data and \emph{XMM-Newton} limits. Also shown is a
single-zone time-dependent model for the continuous injection of
electrons and protons over an assumed age of \snr\ of $2.5$\,kyr
\citep[e.g.][]{1640:Funk}. High-energy electrons produce synchrotron
and IC \g-ray emission in interactions with magnetic and radiation
fields, respectively. High-energy protons produce $\pi^0$-decay \g-ray
emission in interactions with material in the SNR shell. The broadband
SED can be explained in this scenario with a reasonable choice of
input parameters. The leptonic component can be constrained by the
observed synchrotron spectrum from radio to X-rays. In this model
calculation, a magnetic field of $B = 35\,\mu$G, maximum electron
energy of $E_{c,e} = 10$\,TeV and electron spectral index of $\Gamma_e
= 2.0$ is required to reproduce the radio spectrum and to not violate
the X-ray limit. The target radiation fields have been chosen based on
\citet{Lemiere09}, with a dust component that has been increased to
account for the five times higher radiation field energy density in
the northern part of the shell. It is clear from Figure~\ref{fig:snr}
that the predicted IC emission is at least two orders of magnitude
below the observed \g-ray emission for an assumed electron-to-proton
(e/p) ratio of $10^{-2}$. Furthermore, the smooth connection of the HE
and VHE \g-ray spectrum cannot be explained. A considerably higher e/p
ratio of $\simeq0.1$ (and lower magnetic field of $B \simeq 10\,\mu$G)
is required to reach the TeV flux. Even in this case, the IC spectral
shape and maximum energy is not supported by the VHE \g-ray
spectrum. In dense environments, Bremsstrahlung can significantly
contribute to the non-thermal emission. Densities as high as
500\,cm$^{-3}$ and e/p ratios of 0.1 are, however, required to reach
the flux observed by H.E.S.S.

In a hadronic scenario, a total energy transferred into protons of
$W_p = 2.5 \times 10^{50}$\,erg, maximum proton energy $E_{c,p} =
50$\,TeV and spectral index of $\Gamma_p = 2.2$ as well as an average
ambient density $\bar{n}_H = 150$\,cm$^{-3}$, is required to reproduce
the GeV -- TeV spectrum. The measured TeV flux coupled with the large
estimated distance of $\sim10$\,kpc would imply that \hessj\ is the
most luminous Galactic VHE \g-ray SNR detected so far
($L_{>1\mathrm{TeV}} \simeq 4.6\times 10^{35} (d/10\,{\rm
  kpc})^2$\,erg\,s$^{-1}$). The TeV luminosity is therefore about one
order of magnitude higher than that of the W51C SNR
\citep{MAGIC:W51C}. Due to the harder \g-ray spectral index, \hessj\
has a total \g-ray luminosity comparable to W51C. The product of total
energy in interacting protons and mean ambient density of $W_p
\bar{n}_H \simeq 4 \times 10^{52} (d/10\,{\rm kpc})^2$\,erg\,cm$^{-3}$
requires a considerable amount of SN kinetic energy that is
transferred to high-energy protons and/or a high average density of
the target material as motivated before. With the gas densities
estimated above, a very large energy in protons is needed to reach the
measured GeV and TeV flux. This implies that either the SN explosion
was as energetic as $E_{SN} \simeq 4 \times 10^{51} (d/10\,{\rm
  kpc})^2$\,erg (assuming a canonical 10\% of SN explosion energy is
channeled into cosmic rays) and/or that the fraction of $E_{SN}$
transferred into relativistic protons is significantly larger than the
canonical 10\%, i.e. up to $\sim40\,(d/10\,{\rm kpc})^2$\% for a
typical $E_{SN} = 10^{51}$\,erg. Note that this estimate can be even
higher, as only the northern half of the SNR shell seems to be
illuminated by cosmic rays.

\section{Conclusions and Outlook}

The detailed H.E.S.S. results presented in this work show that the VHE
\g-ray emission from \hessj\ significantly overlaps with the
north-western part of the SNR shell of \snr. Moreover, the VHE \g-ray
spectrum smoothly connects with the Fermi spectrum and has a
high-energy cut-off that implies that particles with tens of TeV
energies are present in the acceleration region. The TeV morphology,
new radio measurements and the overall \g-ray spectrum are hard to
explain in a scenario where most of the non-thermal emission is coming
from the PWN. The broadband SED and morphology of the non-thermal
emission from \hessj\ can be better explained in a scenario where
protons are accelerated in the shell of \snr\ and interact with dense
gas associated with the G338.4+0.1 HII complex. In this case, the
product of total energy in interacting protons and mean ambient
density $W_p \bar{n}_H \sim 4\times 10^{52} (d/10\,{\rm
  kpc})^2$\,erg\,cm$^{-3}$ required to explain the flux measured by
Fermi and H.E.S.S. is comparable to the \g-ray-emitting SNR W51C,
although the TeV luminosity of \hessj\ is an order of magnitude
higher. In this picture, the non-detection of thermal X-rays is
consistent with the large distance to \snr\ and the high column
density along the line of sight. High resolution and high sensitivity
molecular line observations in this region are required to locate the
dense gas that might act as target material and to put limits on the
explosion energy of \snr. The future Cherenkov Telescope Array with
its much better angular resolution and sensitivity is needed to
further resolve the VHE \g-ray emission region(s) of \hessj\ and to
distinguish the contribution from the SNR shell and the PWN in \snr.

\section*{Acknowledgements}
The support of the Namibian authorities and of the University of
Namibia in facilitating the construction and operation of H.E.S.S. is
gratefully acknowledged, as is the support by the German Ministry for
Education and Research (BMBF), the Max Planck Society, the French
Ministry for Research, the CNRS-IN2P3 and the Astroparticle
Interdisciplinary Programme of the CNRS, the U.K. Science and
Technology Facilities Council (STFC), the IPNP of the Charles
University, the Czech Science Foundation, the Polish Ministry of
Science and Higher Education, the South African Department of Science
and Technology and National Research Foundation, and by the University
of Namibia. We appreciate the excellent work of the technical support
staff in Berlin, Durham, Hamburg, Heidelberg, Palaiseau, Paris,
Saclay, and in Namibia in the construction and operation of the
equipment. S.O. acknowledges the support of the Humboldt foundation by
a Feodor-Lynen research fellowship. We are also grateful to Gabriela
Castelletti, who kindly provided the 610\,MHz map and Patrick Slane
for the PWN model curve. The authors would also like to thank the
anonymous referee for her/his detailed and constructive comments,
which significantly improved the quality of the paper. This work is
based in part on observations made with the \emph{Spitzer} Space
Telescope, which is operated by the Jet Propulsion Laboratory,
California Institute of Technology under a contract with NASA.

\bibliographystyle{mn2e_williams}
\bibliography{HESSJ1640}
\label{lastpage}

\end{document}